\documentclass[prl,aps,twocolumn,superscriptaddress,showpacs]{revtex4}
\usepackage{amsmath,amssymb,graphicx}
\usepackage{pxfonts}
\usepackage{graphicx}
\begin{document}

\title{Excitonic switches at 100 K temperatures}

\author{G. Grosso}
\author{J. Graves}
\author{A.T. Hammack}
\author{A.A. High}
\author{L.V. Butov}
\affiliation{Department of Physics, University of California at San
Diego, La Jolla, CA 92093-0319}

\author{M. Hanson}
\author{A.C. Gossard}
\affiliation{Materials Department, University of California at Santa
Barbara, Santa Barbara, California 93106-5050}

\begin{abstract}

\end{abstract}


\date{\today}

\maketitle

{\bf Photonic and optoelectronic devices may offer the opportunity to realize efficient signal processing at speeds higher than in conventional electronic devices. Switches form the building blocks for circuits and fast photonic switches have been realized \cite{Liu04,Xu05,Green07,Jiang05,LiuJ08,Chen08}. Recently, proof of principle of exciton optoelectronic devices was demonstrated \cite{High07,High08}. Potential advantages of excitonic devices include high operation and interconnection speed, small dimensions, and the opportunity to combine many elements into integrated circuits. Here, we demonstrate experimental proof of principle for the operation of excitonic switching devices at temperatures around 100 K. The devices are based on an AlAs/GaAs coupled quantum well structure and include the exciton optoelectronic transistor (EXOT), the excitonic bridge modulator (EXBM), and the excitonic pinch-off modulator (EXPOM). This is a two orders of magnitude increase in the operation temperature compared to the earlier devices, where operation was demonstrated at 1.5 K \cite{High07,High08}.}

The operation principle of excitonic devices is based on the control of exciton fluxes by electrode voltages \cite{High07,High08}. The excitonic devices may have photonic or excitonic inputs and outputs. In the former case, photons transform into excitons at the input and excitons transform into photons at the output \cite{High07}. In the latter case, excitons arrive to the input from (depart from the output to) another excitonic device \cite{High08}. Potential advantages of excitonic devices are briefly discussed below.

{\it High interconnection speed}. Efficient signal communication uses photons. Conventional signal processing, however, uses an optically inactive medium, electrons. An interconnection between electronic signal processing and optical communication causes a delay, thus slowing down the operation speed \cite{Miller00}. In contrast, excitons form a medium that can be used for signal processing and, at the same time, directly linked to optical communication. Therefore, the delay between signal processing and optical communication is effectively eliminated in excitonic devices. This can provide a significant advantage in applications where the interconnection speed is important. The switching time combined with the interconnection time for the first proof-of-principle excitonic transistor was $\sim 0.2$ ns \cite{High07}, close to ultrafast photonic modulators.

{\it Compactness and scalability}. Photonic devices demonstrate fast operation speeds; however, the achievement of photonic signal modulation by a control gate usually requires large device dimensions that make the creation of compact circuits with many elements challenging \cite{Wakita98}. The smallest achieved dimensions for Mach-Zehnder modulators are $\sim 100 \mu$m \cite{Liu04,Xu05,Green07,Jiang05,LiuJ08,Chen08}. In comparison, excitonic transistors have an architecture and operation principle similar to electronic field effect transistors (FET). Therefore, excitonic circuits can potentially be similarly compact and can include as many elements as electronic circuits. The diffraction limit for the dimensions of excitonic devices is given by the exciton de Broglie wavelength, much smaller than the diffraction limit for the dimensions of photonic devices given by the wavelength of light (the exciton thermal de Broglie wavelength scales with temperature $\propto T^{-1/2}$ and is $\sim 10$ nm at room temperature). In the first proof-of-principle excitonic transistor the distance between the source and drain was $3 \mu$m \cite{High07} (limited by the resolution of the lithography used for the sample processing), considerably smaller than the dimensions of photonic devices. The proof-of-principle for the integration of excitonic transistors into circuits was demonstrated in \cite{High08}.

\begin{figure}
\includegraphics[width=8.5cm]{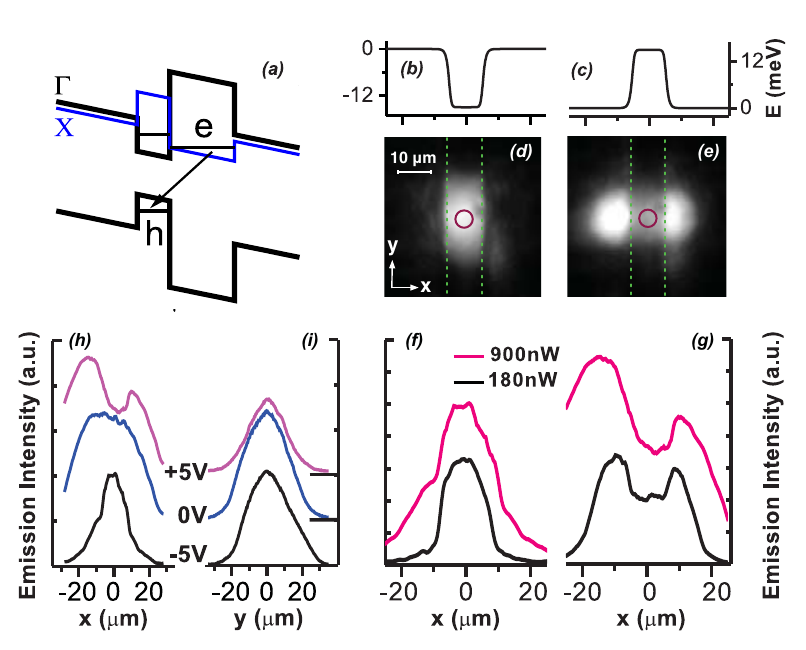}
\caption{{\bf Indirect excitons in electrostatic channels and anti-channels.} (a) Schematic band diagram for AlAs/GaAs CQW. $e$, electron; $h$, hole. Calculated exciton energy profiles for (b) the channel at $V=-5$ V and (c) anti-channel at $V=+5$ V produced by voltage $V$ applied to a linear $10 \mu$m wide electrode. Image of the exciton cloud (d) confined in the channel and (e) spreading from the anti-channel for $P_{ex}=180$ nW. The electrode edges are indicated by the dashed lines. The circle shows the position of the laser excitation. Emission intensity profiles in the (f) channel and (g) anti-channel for $P_{ex}=180$ and 900 nW. Emission intensity profiles (h) across and (i) along the electrode for $V=-5, 0$, and $+5$ V at $P_{ex}=900$ nW. $T=6$ K for all the data.}
\end{figure}

The major challenges for excitonic devices are finite exciton lifetime and finite exciton binding energy. In a regular direct-gap semiconductor, the exciton lifetime is typically less than a nanosecond, allowing the exciton to travel only a small distance before it recombines. This problem is addressed by using indirect excitons composed of electrons and holes in separated layers \cite{High07,High08}. The lifetime of indirect excitons exceeds by orders of magnitude the lifetime of regular excitons and increases exponentially with the separation between the layers $d$ and the height of the separating barrier. Within their lifetime, indirect excitons can travel over large distances \cite{Hagn95,Butov98,Larionov00,Butov02,Voros05,Gartner06,Ivanov06},
for which devices can be readily patterned.

A finite exciton binding energy $E_X$ limits the operation temperature: Excitons exist in the temperature range roughly below $E_X/k_B$ \cite{Chemla84} ($k_B$ is the Boltzmann constant). The proof of principle of excitonic devices was done at 1.5 K in a GaAs/Al$_{0.33}$Ga$_{0.67}$As coupled quantum well structure (CQW) with $d = 12$ nm where $E_X/k_B \sim 40$ K \cite{High07,High08}. Increasing the operation temperature requires different materials and architectures. The goal is to create efficient excitonic devices that are operational at room temperature. Note that materials with a large $E_X$ include wide-bandgap semiconductors \cite{Makino05} and organic materials \cite{Segal07}.

Here, we report on proof of principle for operation of excitonic switches at $T \sim 100$ K. This is a two orders of magnitude increase in the operation temperature compared to the earlier devices, where operation was demonstrated at 1.5 K \cite{High07,High08}. It shows that the operation of excitonic devices is not limited to liquid Helium temperatures. The devices are based on AlAs/GaAs CQW (Fig. 1a). A small $d \sim 3$ nm results in a high binding energy of indirect excitons in such CQW $E_X/k_B \sim 100$ K \cite{Zrenner92}, permitting device operation at elevated temperatures.

\begin{figure}
\includegraphics[width=8.5cm]{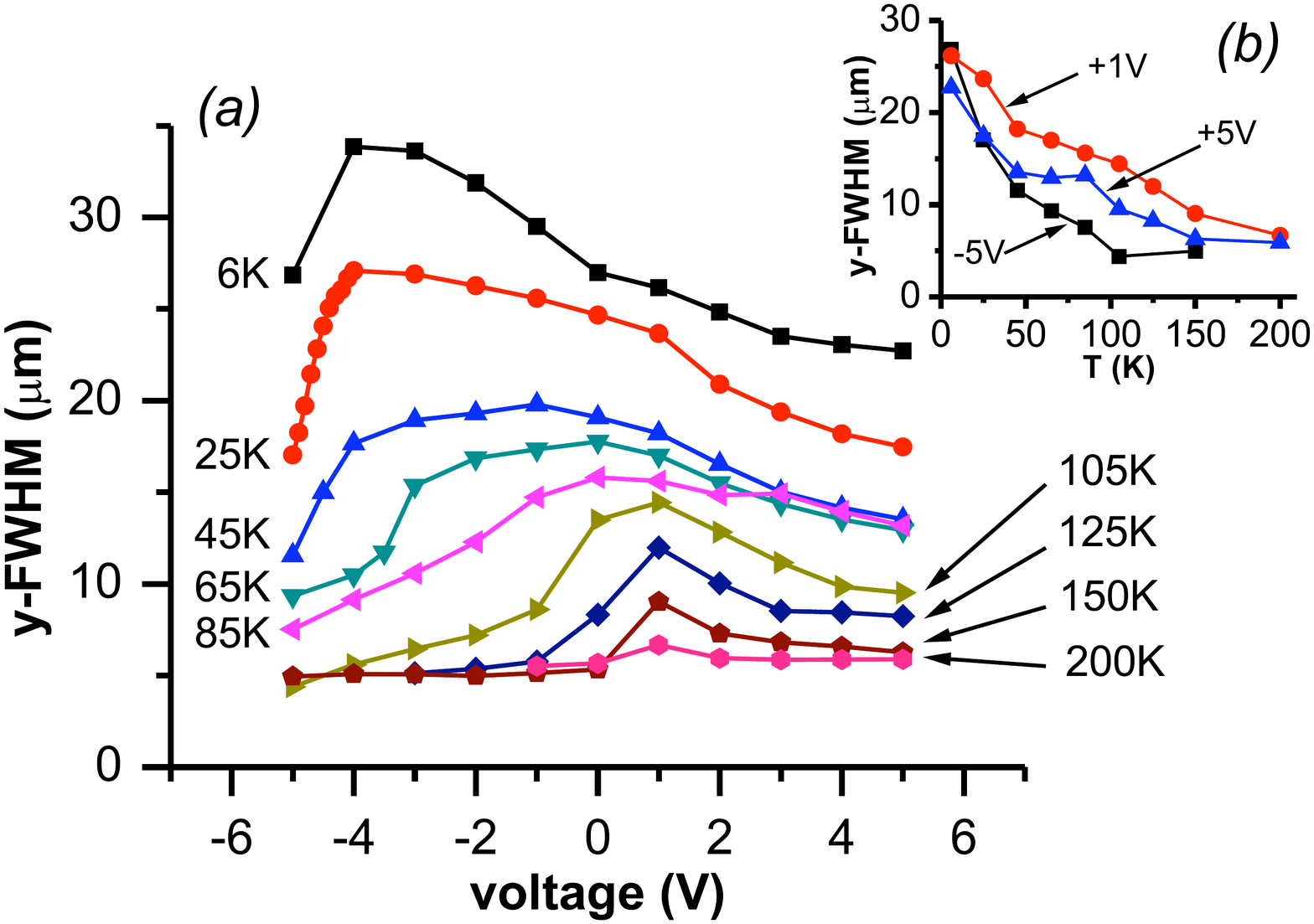}
\caption{{\bf Transport length of indirect excitons in the AlAs/GaAs CQW.} The size (FWHM) of the exciton cloud along the linear electrode (a) vs electrode voltage $V$ for different temperatures and (b) vs temperature for different $V$ at $P_{ex}=900$ nW.}
\end{figure}

We first study exciton transport along and across a $10 \mu$m wide electrode. At zero applied voltage $V$, indirect excitons form the lowest energy state in this structure \cite{Zrenner92}. Application of a positive (negative) $V$ increases (lowers) the exciton energy by $edF_z$ ($F_z$ is the electric field in the z-direction) \cite{Miller85}. The linear electrode creates a channel for indirect excitons at negative $V$ (Fig. 1b)and anti-channel at positive $V$ (Fig. 1c). Figures 1d and 1f show the exciton cloud confinement in the channel. Excitons spread beyond the channel at higher densities (Fig. 1f) due to the screening of the confining potential by the excitons \cite{Remeika09}. Figures 1e and 1g show the exciton cloud spreading away from the anti-channel. The exciton emission reaches maximum intensity a few microns away from the anti-channel (the emission profiles in Figs. 1f-i are corrected for the electrode absorption). Both the increased exciton density away from the potential energy hill of the anti-channel and the increased occupation of the optically active exciton states with momenta $k \le k_0 \approx E_g \sqrt{\varepsilon} / (\hbar c)$ ($E_g$ is the bandgap, $\varepsilon$ is the dielectric constant) due to exciton relaxation to $k \le k_0$ states away from the excitation spot at the potential energy hill contribute to the observed intensity increase beyond the anti-channel. The latter mechanism was suggested as an explanation of the exciton rings in \cite{Butov02,Ivanov06}. Figure 1h shows the transition from a channel to anti-channel with increasing $V$. The confining potential of the channel or potential energy hill of the anti-channel strongly affect the exciton transport in the $x$-direction and relatively weakly affect it in the $y$-direction (Fig. 1h,i). The latter presents the characteristic transport length $l_X$ of indirect excitons in the structure.

Figure 1 shows that $l_X$ is large enough so that excitons can travel over distances exceeding the dimensions of excitonic devices. $l_X$ reduces with increasing temperature (Fig. 2a,b). At high temperatures, the largest $l_X$ is observed around $V \sim 1$ V (Fig. 2a). These data are consistent with earlier data on exciton lifetime $\tau$ in AlAs/GaAs CQW: $\tau$ reduces (i) at positive $V$ due to approaching the direct regime, (ii) at negative $V$ due to the escape of electrons and holes from the CQW through the reduced barrier, and (iii) at high temperatures due to increased nonradiative recombination \cite{Butov94,Butov98}, and the reduction of $\tau$ results in the reduction of $l_X \sim \sqrt{D\tau}$ ($D$ is the exciton diffusion coefficient).

\begin{figure}
\includegraphics[width=8.5cm]{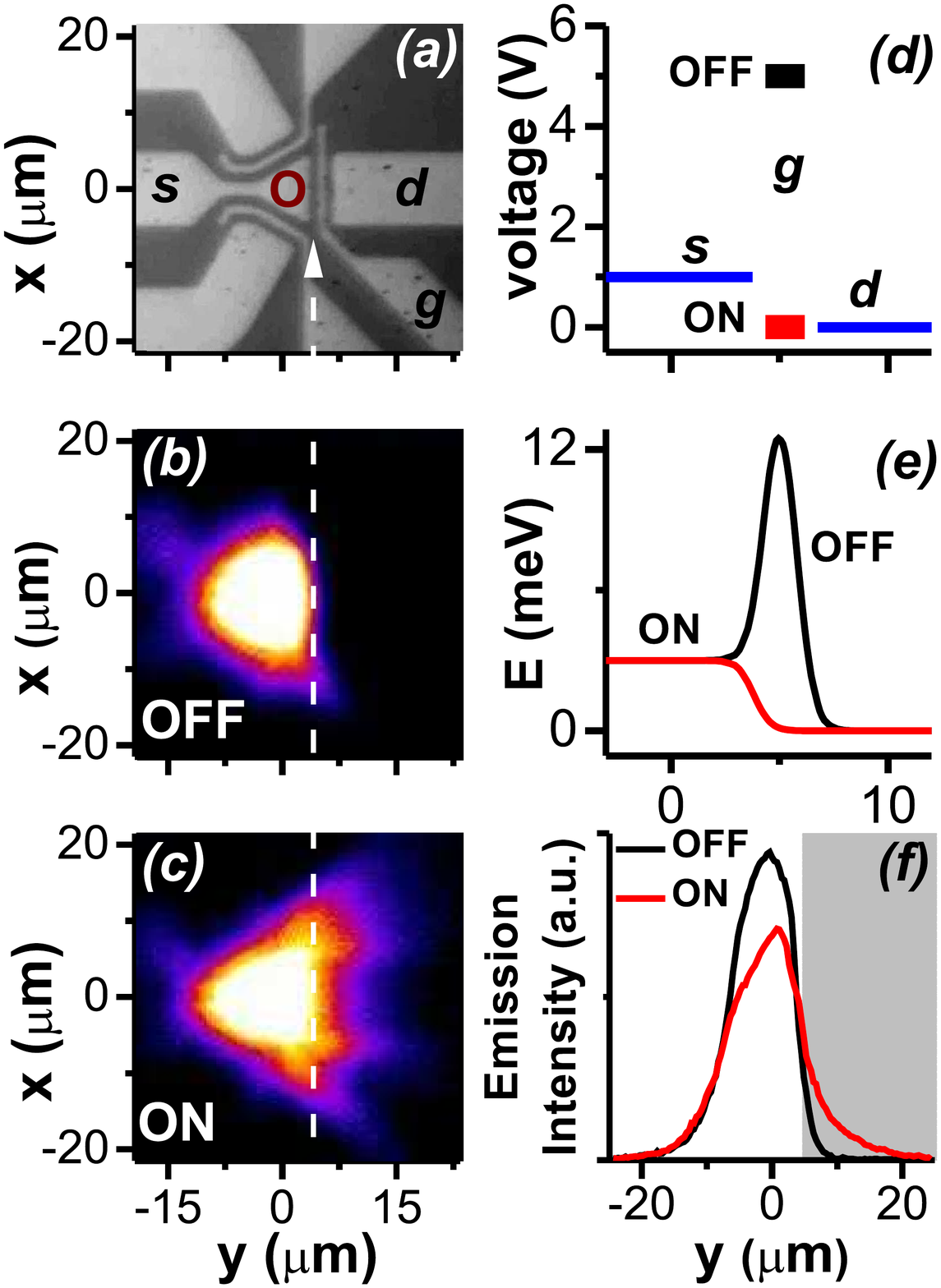}
\caption{{\bf EXOT operation at 85K.} (a) Electrode pattern. The circle shows the position of the laser excitation. The exciton flux from source electrode $s$ to drain electrode $d$ is controlled by a gate electrode $g$, which position is indicated by dashed line. (b,c) Images of the EXOT emission in on and off states at $T=85$ K, $P_{ex}=450$ nW. (d) Voltages and (e) estimated single-exciton energy profiles in on and off states.
$V_s=1, V_d=0, V_g=0$ in on and 5 V in off state and the voltage applied to the rest of the electrodes is 5 V. (f) Emission intensity along the exciton flux for off (red line) and on (black line) states [correspond to the false-color images in (b,c)]. Gray area indicates the drain region.}
\end{figure}

Figure 3 presents the proof of principle for the operation of the exciton optoelectronic transistor (EXOT) at 85 K. The operation principle of the EXOT is similar to that of an electronic FET \cite{High07,High08}. The EXOT is a three-terminal device in which the exciton flux between two electrodes is controlled by a voltage applied to the third electrode. Here, the EXOT operates in switching mode. The excitons are excited at the source (input) and travel to the drain (output) because of the potential energy gradient $\sim ed(F_{zd} - F_{zs}) \propto (V_d - V_s)$ created by the source voltage $V_s$ and drain voltage $V_d$. The exciton flux from source to drain is controlled by a gate voltage $V_g$, which creates (removes) a barrier in the region of the gate electrode when the EXOT is off (on), see Fig. 3d,e. The emission images for the EXOT in both the off state and on state are shown in Fig. 3b,c. The on/off ratio of the signal integrated over EXOT output (shaded region) reaches 5 (Fig. 3f).

\begin{figure}
\includegraphics[width=8.5cm]{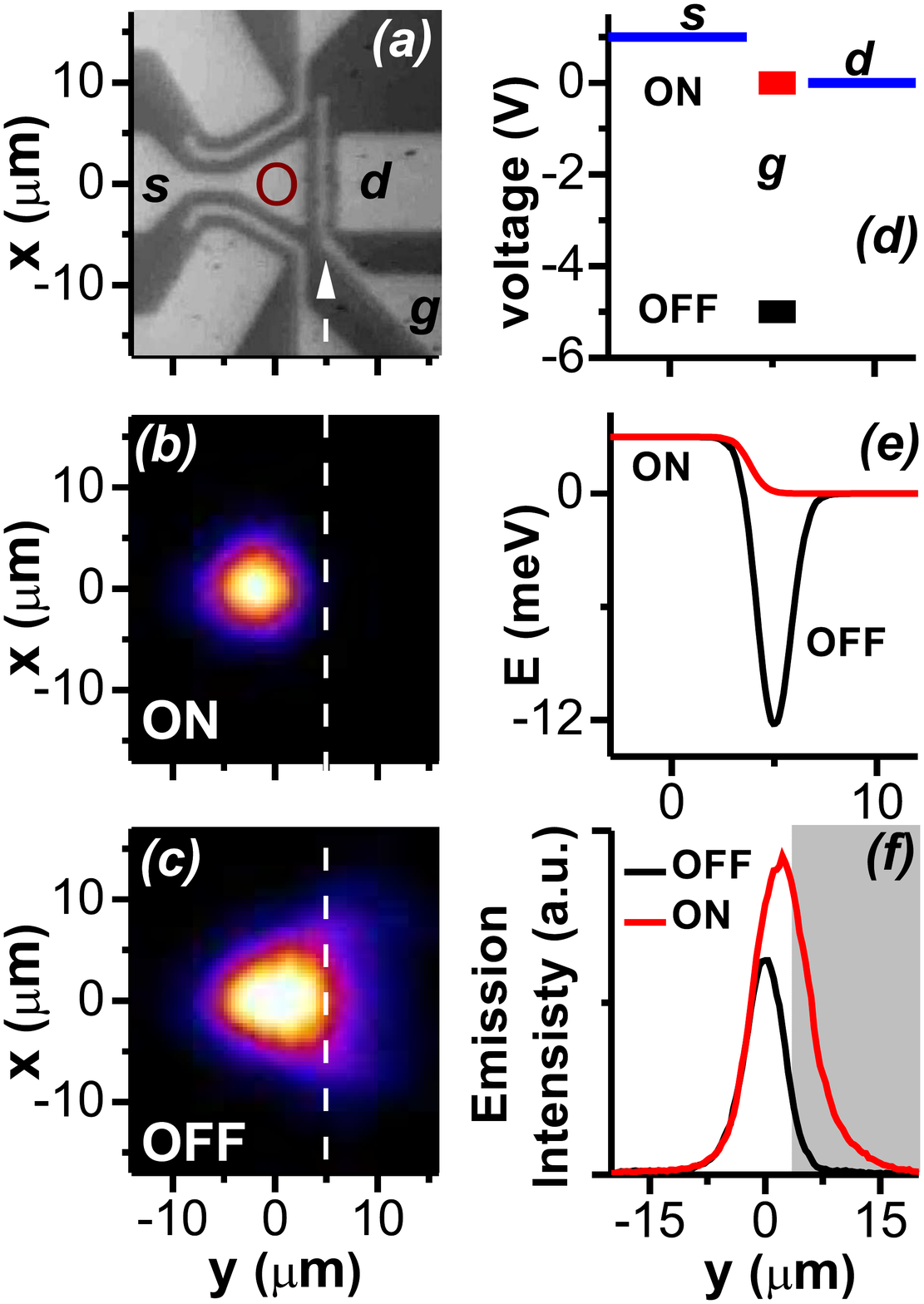}
\caption{{\bf EXBM operation at 125K.} (a) Electrode pattern. The circle shows the position of the laser excitation. The exciton flux from source electrode $s$ to drain electrode $d$ is controlled by a gate electrode $g$, which position is indicated by dashed line. (b,c) Images of the EXBM emission in on and off states at $T=125$ K, $P_{ex}=4.5 \mu$W. (d) Voltages and (e) estimated single-exciton energy profiles in on and off states. $V_s=1, V_d=0, V_g=0$ in on and $-5$ V in off state and the voltage applied to the rest of the electrodes is $-5$ V. (f) Emission intensity along the exciton flux for off (red line) and on (black line) states [correspond to the false-color images in (b,c)]. Gray area indicates the drain region.}
\end{figure}

Exciton flux modulations can also be achieved by other configurations of electrode voltages. Figure 4 presents the proof of principle for the operation of the excitonic bridge modulator (EXBM) at 125 K. In this device, the exciton flux from source to drain is terminated in the off state by a considerable reduction of the exciton lifetime in the gate region. This is achieved by applying a large negative voltage on the gate electrode, which increases the carrier escape out of the CQW in the $z$-direction in the gate region \cite{Butov94}. In turn, the exciton flux can flow from source to drain when the gate electrode provides a bridge between the source and drain with a long exciton lifetime. The on/off ratio of the signal integrated over EXBM output (shaded region) reaches 9 (Fig. 4f).

\begin{figure}
\includegraphics[width=8.5cm]{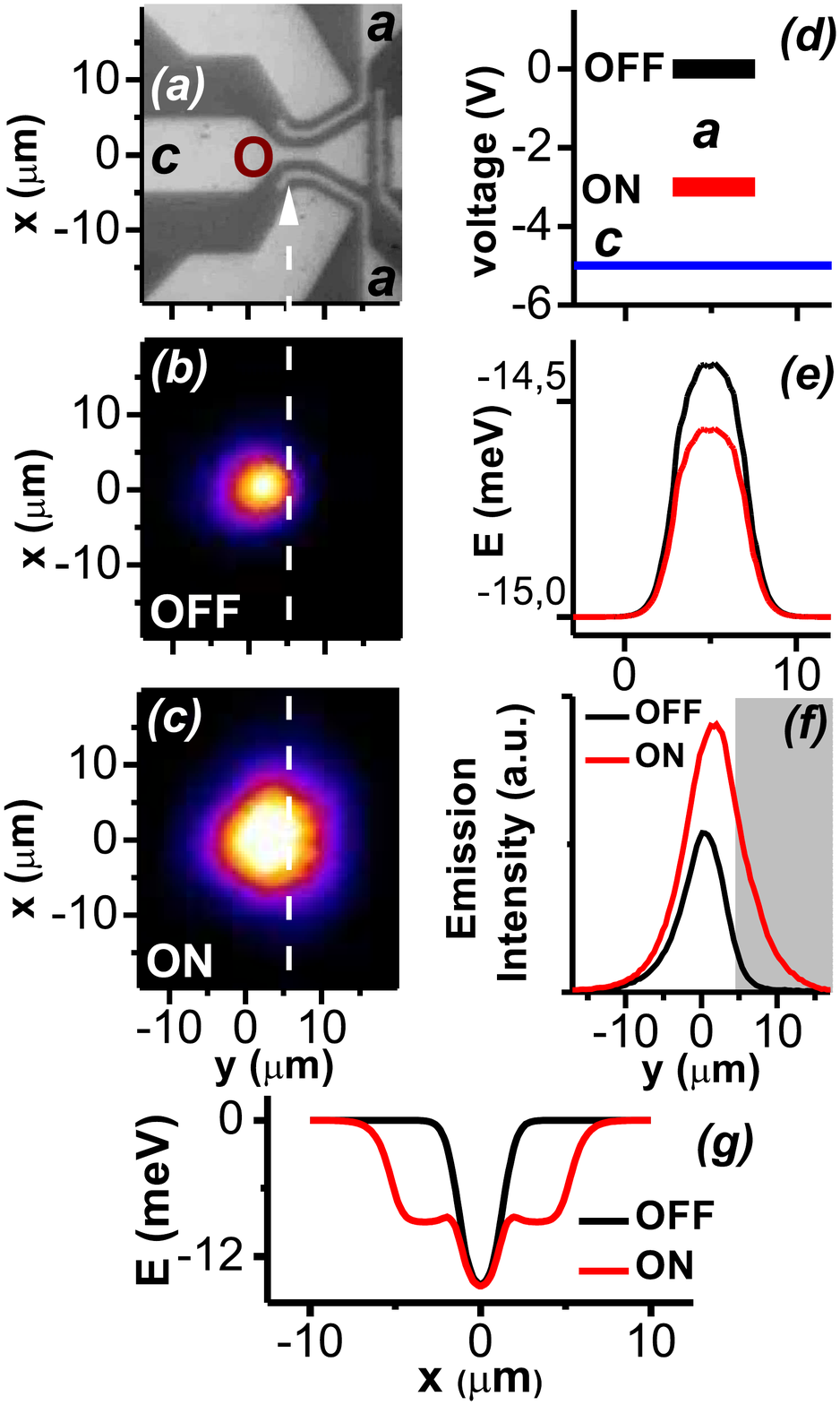}
\caption{{\bf EXPOM operation at 85K.} (a) Electrode pattern. The circle shows the position of the laser excitation. The exciton flux through the constriction in channel electrode $c$, which position is indicated by dashed line, is controlled by the voltage on adjacent electrodes $a$.
(b,c) Images of the EXPOM emission in on and off states at $T=85$ K, $P_{ex}=6.3 \mu$W. (d) Voltages and estimated single-exciton energy profiles in on and off states in (g) $x$ and (e) $y$ directions. $V_c=-5, V_a=-3$ in on and $0$ V in off state and the voltage applied to the rest of the electrodes is 0 V. (f) Emission intensity along the exciton flux for off (red line) and on (black line) states [correspond to the false-color images in (b,c)]. Gray area indicates the region right of the channel.}
\end{figure}

Figure 5 presents the proof of principle for the operation of the excitonic pinch-off modulator (EXPOM) at 85 K. In this device, the exciton flux through the $2 \mu$m constriction in the electrode -- the channel -- is controlled by varying the voltage on the laterally adjacent electrodes $V_a$. Increasing $V_a$ narrows the channel and also increases the exciton energy in the channel (Fig. 5e,g), reducing the exciton flux through it. The EXPOM is similar to split gate electronic devices. The on/off ratio of the signal integrated over the area right of the channel (shaded region) reaches 12 (Fig. 5f).

At high temperatures $\sim 100 \-- 150$ K, the exciton transport length becomes small compared to the device dimensions (Fig. 2), $E_X/k_B$ become comparable to the temperature, and the on/off ratio for the switches in the AlAs/GaAs CQW vanishes. The optimization of excitonic devices is the subject of on-going research.

This work is supported by ARO and NSF.

\vskip 1cm

{\bf Methods}
The structure was grown by molecular beam epitaxy. An $n^+$-GaAs layer with $n_{Si}=10^{18}$ cm$^{-3}$ serves as a homogeneous bottom electrode. The top electrodes were fabricated by depositing a semitransparent layer of Ti (2 nm) and Pt (8 nm). The 2.5 nm GaAs QW and 4 nm AlAs QW were positioned $0.1 \mu$m above the $n^+$-GaAs layer within an undoped $1 \mu$m thick Al$_{0.45}$Ga$_{0.55}$As layer. Positioning the CQW closer to the homogeneous electrode suppresses the in-plane electric field \cite{Hammack06}, which otherwise can lead to exciton dissociation. Excitons were photoexcited by a 633 nm HeNe laser within a $4 \mu$m FWHM (full width at half maximum) spot. The emission images were taken by a CCD with an interference filter covering the spectral range of the indirect excitons emitting in the vicinity of 690 nm or using a spectrometer with resolution 0.4 meV. The spatial resolution was $1.5 \mu$m. The single-exciton energy profiles were estimated as in \cite{Hammack06}.

\end{document}